# CROSS-LAYER RESOURCE ALLOCATION SCHEME UNDER HETEROGENEOUS CONSTRAINTS FOR NEXT GENERATION HIGH RATE WPAN

Ayman Khalil, Matthieu Crussière and Jean-François Hélard

INSA, 20 avenue des Buttes de Coësmes, 35708 Rennes, France
Institute of Electronics and Telecommunications of Rennes (IETR)
ayman.khalil@insa-rennes.fr

## ABSTRACT

*In the next generation wireless networks, the growing demand for new wireless applications is accompanied with high expectations for better quality of service (QoS) fulfillment especially for multimedia applications. Furthermore, the coexistence of future unlicensed users with existing licensed users is becoming a challenging task in the next generation communication systems to overcome the underutilization of the spectrum. A QoS and interference aware resource allocation is thus of special interest in order to respond to the heterogeneous constraints of the next generation networks. In this work, we address the issue of resource allocation under heterogeneous constraints for unlicensed multi-band ultra-wideband (UWB) systems in the context of Future Home Networks, i.e. the wireless personal area network (WPAN). The problem is first studied analytically using a heterogeneous constrained optimization problem formulation. After studying the characteristics of the optimal solution, we propose a low-complexity suboptimal algorithm based on a cross-layer approach that combines information provided by the PHY and MAC layers. While the PHY layer is responsible for providing the channel quality of the unlicensed UWB users as well as their interference power that they cause on licensed users, the MAC layer is responsible for classifying the unlicensed users using a two-class based approach that guarantees for multimedia services a high-priority level compared to other services. Combined in an efficient and simple way, the PHY and MAC information present the key elements of the aimed resource allocation. Simulation results demonstrate that the proposed scheme provides a good tradeoff between the QoS satisfaction of the unlicensed applications with hard QoS requirements and the limitation of the interference affecting the licensed users.*

## 1. INTRODUCTION

While the next generation wireless networks are being driven by a large set of new application requirements, they are promising to provide higher data rates and better quality of service (QoS) achievement especially for multimedia applications. Thereby, as the number of wireless applications is increasing, the coexistence of various types of wireless devices is one of the major challenging issues.

On the other hand, while spectrum resource has become increasingly scarce, spectrum occupancy measurements have shown that most of the assigned radio spectrum is still significantly underutilized. In that context, cognitive radio (CR) and ultra-wideband (UWB) are two recent and exciting technologies that offer new solutions for the spectrum scarcity and spectrum underutilization. CR is based on the spectrum sensing and dynamic spectrum access (DSA) techniques to find available spectrum which can be used by a CR user without causing any harmful interference to licensed users [1]. The UWB approach is however based on an underlay usage of the spectrum obtained under tough power spectral density (PSD) limitations. With this latter solution, even if the transmission mask should protect existing systems, there is interest in finding a flexible method for managing the spectrum access of the secondary users in order to satisfy high-priority multimedia services while reducing the interference they cause on the licensed users.

So far, UWB has been attracting great interest as an appropriate technology for unlicensed next generation short range communications. In 2002, the Federal Communications Commission (FCC) regulated UWB systems by allocating them the 3.1 to 10.6 GHz spectrum for unlicensed use [2], with a power spectral density level restricted to a maximum of -41.3 dBm/MHz. This stringent power limitation should ensure an underlay usage of the spectrum with little effects on other licensed services. One of the techniques proposed for high-rate UWB by the IEEE802.15.3a workgroup is based on orthogonal frequency division multiplexing (OFDM) referred to as MB-OFDM [3]. Since 2005, this MB-OFDM approach has been supported by the WiMedia Alliance that promoted it to the ECMA International standardization body. As a result, on December 2005, ECMA International eventually approved two standards for UWB technology [4] based on the MB-OFDM WiMedia solution: ECMA-368 for high-rate UWB PHY and MAC Standard and ECMA-369 for MAC-PHY Interface for ECMA-368. However, these standardized UWB systems do not integrate any efficient adaptive mechanisms for dynamic spectrum usage and interference mitigation.

In the literature, the resource allocation problem in OFDM systems is addressed in general as an optimization problem where optimal and suboptimal subcarrier and power allocation are proposed using one of the two well-known optimization classes: margin adaptive and rate adaptive [5, 6]. Besides, some studies have been devoted to the resource allocation problem under interference constraints. In [7] for instance, the authors define an optimization allocation problem that maximizes the system throughput (i.e. the sum of the rates of all users) while limiting the interference caused by each unlicensed user. Comparatively, in [8], a suboptimal resource allocation algorithm that minimizes the total interference caused by unlicensed users as well as their transmitted power is presented. In [9], the authors introduce optimal and suboptimal power loading algorithms for an OFDM-based CR system under interference limitations.

In MB-OFDM UWB, few related studies are proposed for spectrum sharing and resource allocation. The authors in [10] focus on the adaptive subcarrier selection and power allocation in OFDM-based UWB systems in a single-user scheme. A multiuser optimal and suboptimal sub-band and power allocation scheme for the multiband UWB systems is also proposed in [11] under limited power constraint.

None of the above mentioned studies takes into consideration the service differentiation and the QoS support for multimedia and real-time applications with the interference constraint. In the perspective of a diversification of the application requirements, this feature has however to be considered in the design of allocation algorithms.

The objective of this paper is to study a novel heterogeneous resource allocation optimization scheme for the MB-OFDM UWB systems under interference and QoS constraints so that we can make a proper tradeoff between the QoS support and the interference level in order to satisfy both the unlicensed multimedia users and the primary users. Eventually, we aim at defining a new allocation scheme for the unlicensed UWB users that takes into account the following three major criteria: (i) the service differentiation issue through the classification of the UWB users in two classes: Hard-QoS (or HQoS) class for multimedia applications, and Soft-QoS (or SQoS) class for data applications; (ii) the channel state information (CSI) through the use of the effective SINR method; and (iii) the limitation of the interference introduced by the unlicensed UWB users on the existing licensed users.

The paper is organized as follows. In section 2, the MB-OFDM UWB system model is introduced. Then, an analytical study is defined in section 3. The appropriate allocation criteria, i.e. the service differentiation, the CSI and the interference power are presented in order to serve the optimization problem. Based on the aspects presented in section 3, the constrained optimization problem is derived in section 4. We propose in section 5 two algorithms for the addressed problem; one optimal algorithm based on an iterative procedure to obtain the optimal sub-band and power allocation, and one low-complexity suboptimal algorithm based on a simple cross-layer approach. Section 6 gives simulation results for the proposed algorithms and provides analysis and comparisons between the performance of the optimal and suboptimal solutions. Besides, we present some analysis on the interference impact on licensed users as well as the performance of the UWB users in terms of power and rate satisfaction in order to check the QoS support of users having strict QoS requirements.

## 2. SYSTEM MODEL: MB-OFDM SOLUTION

The MB-OFDM UWB approach (or equivalently the WiMedia solution) consists in combining OFDM with a multi-banding technique that divides the available band into 14 sub-bands of 528 MHz each, as illustrated in Fig. 1. Five band groups or channels are defined, each consisting of three consecutive sub-bands, except for the fifth one which includes only the last two sub-bands. A WiMedia compatible device should actually make use of only one out of these five defined channels. As mentioned in the introduction, note that the PSD level is set to -41.3 dBm/MHz for UWB systems.

In this system, the MB-OFDM scheme is applied with a total of 128 subcarriers per band, 100 data carriers, 10 guard carriers, 12 pilot and 6 null tones. The OFDM signal can be transmitted on each sub-band using a 128-point inverse fast Fourier transform (IFFT). The pilot tones are used in order to achieve the coherent detection. The constellation applied to the different subcarriers is either a quadrature phase-shift keying (QPSK) for the low data rates or a dual carrier modulation (DCM) for the high data rates. Different data rates from 53.3 to 480 Mbps are obtained with the combined use of forward error correction (FEC), frequency-domain spreading (FDS) and time-domain spreading (TDS), as presented in Table 1. This enables optimum performance under a variety of channel conditions varying the information data rate of the system. The FEC used is a convolutional code with coding rates of 1/3, 1/2, 5/8 and 3/4. Note that in Table 1, a new parameter $\lambda$ is introduced. This parameter will be defined in section 3and used for the exploitation of the CSI at the physical level.

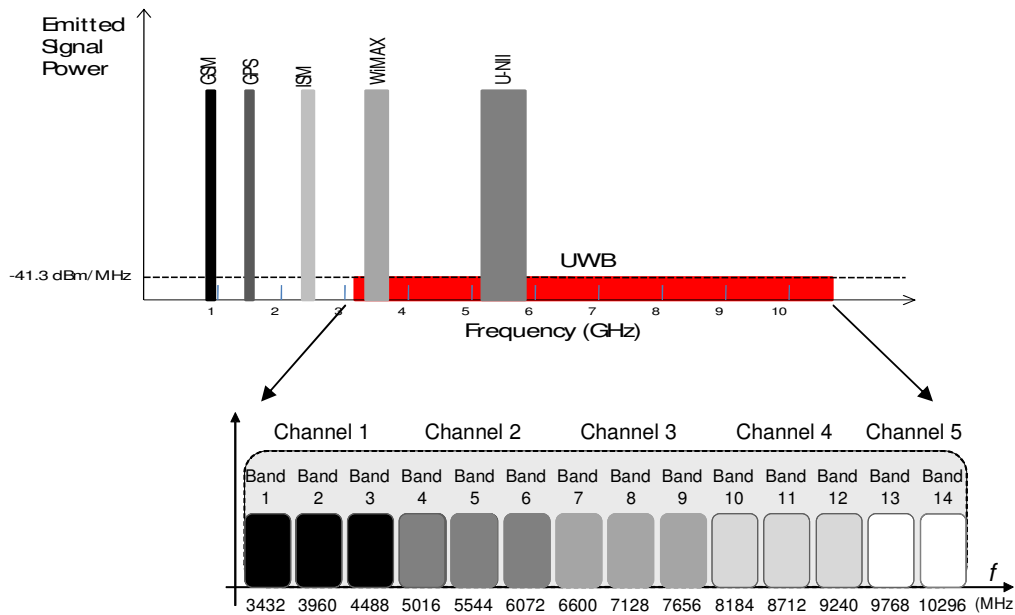

Fig. 1. UWB spectrum and sub-band distribution for WiMedia solution.

Table 1. WiMedia data rates and the associated parameter $\lambda$.

| Data Rate (Mbps) | Modulation | Coding Rate (r) | Frequency Domain Spreading | Time Spreading Factor | Code bits per OFDM symbol | $\lambda$ |
|---|---|---|---|---|---|---|
| 53.3 | QPSK | 1/3 | Yes | 2 | 100 | 1.49 |
| 80 | QPSK | 1/2 | Yes | 2 | 100 | 1.57 |
| 110 | QPSK | 11/32 | No | 2 | 200 | 1.42 |
| 160 | QPSK | 1/2 | No | 2 | 200 | 1.57 |
| 200 | QPSK | 5/8 | No | 2 | 200 | 1.82 |
| 320 | DCM | 1/2 | No | 1 | 200 | 1.85 |
| 400 | DCM | 5/8 | No | 1 | 200 | 1.82 |
| 480 | DCM | 3/4 | No | 1 | 200 | 1.80 |

The coded data is spread by using a time-frequency code (TFC). In all, there are two types of TFCs. The first one is a time-frequency interleaving (TFI), where the coded information is interleaved over three bands and the second one is a fixed frequency interleaving (FFI), where the coded information is transmitted on a single band. TFC allows each user to benefit from frequency diversity over a bandwidth equivalent to the two or three sub-bands of one channel. In addition, to prevent interference between consecutive symbols, a zero padding (ZP) guard interval is added instead of the traditional cyclic prefix (CP) used in the classical OFDM systems. The ZP simply consists in trailing zeros and requires a specific processing at the receiver side to compensate for the lack of cyclic structure in the received signal and hereby make possible a simple OFDM demodulation through FFT computation [12].

From the physical layer point of view, the WiMedia solution offers potential advantages for high-rate UWB applications, such as the signal robustness against channel selectivity and the efficient exploitation of the energy of every signal received within the prefix margin.

Concerning the MAC layer and the medium access, a combination of carrier sense multiple access (CSMA) and time division multiple access (TDMA) is adopted to ensure prioritized schemes for isochronous and asynchronous traffics. To reserve any TDMA access for isochronous and other data transfer, a distributed reservation protocol (DRP) is used. For network scalability, meaning possible less use of extra resources when the number of devices increases, prioritized contention access (PCA) is provided using a CSMA scheme [4].

Channel time is divided into superframes. A superframe is the basic timing structure for frame exchange and it is composed of two major parts, the beacon period (BP) and the data transfer period (DTP). The duration of the superframe is specified as 65536 μs, and the superframe consists of 256 medium access slots (MASs), which are all of equal length, 256 μs.

## 3. CROSS-LAYER INFORMATION

The proposed allocation scheme counts on the collection of information located at two different levels, more precisely the PHY and the MAC levels. In this section, we present the new functionalities of these two layers that should contribute to the optimization problem formulation.

### 3.1. MAC layer Information

#### 3.1.1. Service classification

Since multimedia applications services are key applications in next generation wireless networks, especially in high-rate UWB networks, it is desirable to assign them a high level of priority in any radio access mechanism. A two-level service classification model is proposed in this paper to ensure the prioritization principle and to respond to next generation systems QoS requirements. Consequently, we classify the UWB service types into two classes:

1. Hard-QoS class (HQoS): This class is defined for applications or services that require strong QoS support, more precisely multimedia applications. Voice and video services for instance are non delay-tolerant applications, they have thus strict QoS requirements and they definitely belong to this class.

2. Soft-QoS class (SQoS): This class is dedicated to applications that don't have strict QoS requirements, more precisely non real-time or data applications. Best effort (BE) and file transfer services for instance are delay-tolerant applications, they belong thus to this class.

### 3.1.2. Weight assignment

The defined service classification scheme offers a two-level priority-based model which affects the scheduling decision. Effectively, we assign a class weight to the different users or applications belonging to the two defined classes. A higher weight is thus to be assigned to the service type with strict QoS requirements.

Our weight assignment model is divided into two parts: fixed class weight assignment and dynamic service weight assignment:

- **Fixed class weight**

A constant weight is assigned to the users aiming at accessing the network based on the class to which they belong. This approach is proposed in [16] for IEEE 802.16 where the four defined classes are assigned four weights according to the class priority. Similarly, according to our two-level service classification model, weight 2 is attributed to HQoS class and weight 1 to SQoS class.

- **Dynamic service weight**

Since different services belonging to the same class may have different QoS requirements, we define a dynamic service weight that ensures an additional level of differentiation between users according to their requested data rates. Consequently, a user $k$ is assigned a service weight $s_k$ defined as

$$s_k = 1 + \frac{R_k - R_{\min}}{R_{\max} - R_{\min}} \qquad (1)$$

where $R_k$ is the user $k$ requested data rate, $R_{\min}$ and $R_{\max}$ are respectively the lowest and the highest data rates taken from the WiMedia data rate modes as presented in Table 1. Evidently, this service weight gives advantage to users having high data rate requirements.

Note that if two or more users require the same data rate which results in assigning them the same weight, an additional differentiation level is demanded which is the delay tolerance. For instance, two users belonging to the same class and having the same rate requirements have to be differentiated according to their target delay requirement, so that the user with a lower delay tolerance is considered as a higher priority user.

- **Absolute user weight**

Provided by the MAC layer, the fixed and the dynamic weight definitions ensure an adaptive rate differentiation for the end-users according to their requirements and to the system constraints. Accordingly, the absolute user weight $W$ is the combination of the class weight with the service weight defined as

$$W_k = q_k \times s_k \qquad (2)$$

where $q_k$ is the user $k$ class weight and $s_k$ is its service weight.

## 3.2. PHY layer Information

### 3.2.1. Channel State Information (CSI)

Since the unlicensed UWB users have to learn about the channel conditions to adjust their transmission parameters, useful channel information can be provided to each user by exploiting the CSI. Assuming that the instantaneous SINR for each subcarrier is known by each user, it is possible to evaluate the system level performance in terms of BER by using the effective SINR approach. The basic idea of the effective SINR method is to find a compression function that maps the sequence of varying SINRs to a single value that is correlated with the BER. This new channel approach used in the 3GPP standardization as an effective link to system mapping method [13,14] is useful in representing the quality of a sub-band by a scalar value stated as

$$SINR_{eff} = I^{-1}\left(\frac{1}{N}\sum_{i=1}^{N} I(SINR_i)\right) \qquad (3)$$

where $I(x)$ is called the information measure function, $N$ the number of subcarriers in a sub-band and $SINR_i$ the ratio of signal to interference and noise for the $i$th subcarrier. Referring to the Exponential Effective SINR Mapping (EESM), we use the following expression for $I(x)$

$$I(x) = \exp(-\frac{x}{\lambda}) \qquad (4)$$

where $\lambda$ is a scaling factor that depends on the selected modulation and coding scheme (MCS). In our system, $\lambda$ is computed and evaluated for the eight WiMedia data rate modes as presented in Table 1.

Eventually, the effective SINR writes

$$SINR_{eff} = -\lambda \ln\left[\frac{1}{N}\sum_{i=1}^{N}\exp(-\frac{SINR_i}{\lambda})\right] \qquad (5)$$

In practice, based on the CSI knowledge, each user is capable to compute the effective SINR value in each sub-band by using (5). For instance, in the case of one channel divided into $B = 3$ sub-bands and with $K = 3$ users, the physical layer information is reduced to the knowledge of only $B \times K = 9$ effective SINR values.

### 3.2.2. Interference Power

With the limited power imposed by the FCC to the unlicensed UWB users, we have to be aware in any power allocation scheme to limit the interference that could be caused by the UWB users to the primary users sharing the same spectrum. More precisely, since we are dealing with heterogeneous environment due to different service classes or traffic types, this will absolutely lead to different power level assignments. Our objective is thus to control the power assignment of the different UWB users in order to limit or reduce the interference caused by these users on the primary users.

According to [15], in OFDM systems the interference power caused by a secondary user $k$ assigned a subcarrier $i$ and affecting a primary user $u$ is defined as

$$I_{k,i}^{u} = P_{k,i} I_i \qquad (6)$$

$$\text{where } I_i = \int_{(n_i^u - \frac{1}{2})\Delta f}^{(n_i^u + \frac{1}{2})\Delta f} \Phi_i(f)df \tag{7}$$

where $P_{k,i}$ is the power allocated to user $k$ in subcarrier $i$, $n_i^u$ the spectral distance between subcarrier $i$ and the center of primary user $u$ band, $\Delta f$ the bandwidth of primary user $u$ band and $\Phi_i(f)$ the spectral pulse shape of subcarrier $i$.

We extend these formulas to the MB-OFDM systems to obtain

$$I_{k,b}^u = P_{k,b} I_b \tag{8}$$

$$\text{where } I_b = \sum_{i=1}^{N} \int_{(n_i^u - \frac{1}{2})\Delta f}^{(n_i^u + \frac{1}{2})\Delta f} \Phi_i(f)df \tag{9}$$

where $P_{k,b}$ is the power allocated to user $k$ in sub-band $b$ and $N$ the number of subcarriers in one sub-band.

## 4. PROBLEM FORMULATION

In order to address the resource allocation matter in a heterogeneous context under QoS and interference requirements, we first study it analytically by deriving a constrained optimization problem. We consider a system consisting of $U$ primary users and $K$ UWB users where the first $K_h$ users are HQoS users and the remaining $K-K_h$ are SQoS users. The rate of a user $k$ in sub-band $b$ is defined as

$$r_{k,b} = \log_2(1 + P_{k,b} E_{k,b}) \tag{10}$$

where $E_{k,b}$ is the effective SINR of user $k$ in sub-band $b$. The objective is to find a joint sub-band and power allocation scheme for the UWB users in a fair way that maximizes the total data rate of the $K-K_h$ SQoS users while respecting the following conditions:

- maintaining a certain level of transmission rate for the $K_h$ HQoS users,
- limiting or reducing the interference power caused by the UWB users on the primary users,
- and respecting the total transmission power $P_T$ constraint of the UWB systems.

The problem can be formulated as

$$\begin{aligned}
\max_{S_k, P_k} \quad & \sum_{k=K_h+1}^{K} \sum_{b \in S_k} r_{k,b} \\
\text{subject to} \quad & \sum_{b \in S_k} r_{k,b} \geq R_k, \quad k = 1,...,K_h \\
& \sum_{k=1}^{K} \sum_{b=1}^{B} I_{k,b}^u \leq I_u^{th}, \quad u = 1,...,U \\
& \sum_{k=1}^{K} \sum_{b=1}^{B} P_{k,b} \leq P_T
\end{aligned} \tag{11}$$

where $B$ is the total number of sub-bands, $R_k$ the HQoS user $k$ required data rate, $I_u^{th}$ the power interference threshold defined by the primary user $u$, $S_k$ the set of sub-bands assigned to user $k$. In our case, $S_1$, $S_2$,.., $S_k$ are disjoint and each user is assigned one sub-band during one time interval. This problem is a mixed integer linear programming problem since $S_k$ are integer variables [17]. Consequently, the problem is classified as NP-hard. A method that makes the problem solvable is to relax the constraint that each sub-band is assigned to one user only. The idea is to allow the users to time-share each sub-band by defining a new parameter $\rho_{k,b}$, which represents the time-sharing factor for user $k$ of sub-band $b$. The optimization problem can be reformulated as

$$\max_{P_{k,b}, \rho_{k,b}} \quad \sum_{k=K_h+1}^{K} \sum_{b=1}^{B} \rho_{k,b} \log_2(1+\frac{P_{k,b}E_{k,b}}{\rho_{k,b}})$$

$$subject \ to \ \sum_{b=1}^{B} \rho_{k,b} \log_2(1+\frac{P_{k,b}E_{k,b}}{\rho_{k,b}}) \geq R_k, \quad k=1,.....K_h$$

$$\sum_{k=1}^{K} \rho_{k,b} = 1, \quad \forall b \quad 0 \leq \rho_{k,b} \leq 1 \quad \forall k,b \qquad (12)$$

$$\sum_{k=1}^{K} \sum_{b=1}^{B} I_{k,b}^{u} \leq I_u^{th}, \quad u=1,...,U$$

$$\sum_{k=1}^{K} \sum_{b=1}^{B} P_{k,b} \leq P_T$$

The problem in (12) is a convex maximization problem. Using standard optimization techniques, we obtain the Lagrangian

$$L = \sum_{k=K_h+1}^{K} \sum_{b=1}^{B} \rho_{k,b} \log_2(1+\frac{P_{k,b}E_{k,b}}{\rho_{k,b}}) \ + \sum_{k=1}^{K_h} \alpha_k \ (\sum_{b=1}^{B} \rho_{k,b} \log_2(1+\frac{P_{k,b}E_{k,b}}{\rho_{k,b}}) - R_k) +$$

$$\sum_{b=1}^{B} \beta_b (1 - \sum_{k=1}^{K} \rho_{k,b} \ ) \ + \gamma(I_u^{th} - \sum_{k=1}^{K} \sum_{b=1}^{B} I_{k,b}^{u}) + \theta(P_T - \sum_{k=1}^{K} \sum_{b=1}^{B} P_{k,b}) \qquad (13)$$

### Sub-band Allocation

Let $\rho_{k,b}^*$ be the optimal solution. After differentiating (13) with respect to $\rho_{k,b}$ by KKT optimality condition [18], we obtain

$$\alpha_k \left[ \log_2(\frac{\alpha_k E_{k,b}}{\gamma \ln 2}) - \frac{1}{\ln 2}(1 - \frac{\gamma \ln 2}{\alpha_k E_{k,b}}) \right] - \beta_b = 0, \qquad for \ k=1,..,K_h$$

$$\log_2(\frac{E_{k,b}}{\gamma \ln 2}) - \frac{1}{\ln 2}(1 - \frac{\gamma \ln 2}{E_{k,b}}) - \beta_b = 0, \qquad for \ k=K_h+1,..,K$$

$$(14)$$

Since $\rho_{k,b}^*$ should satisfy the following KKT condition

$$\frac{\partial L}{\partial \rho_{k,b}^*} = \begin{cases} >0, & \rho_{k,b}^* = 1 \\ =0, & 0 < \rho_{k,b}^* < 1 \\ <0, & \rho_{k,b}^* = 0 \end{cases} \qquad (15)$$

Substituting (14) into (15), we get

$$\rho_{k,b}^* = \begin{cases} 1, & H_{k,b} > \beta_b \\ 0, & H_{k,b} < \beta_b \end{cases} \qquad (16)$$

where $H_{k,b}$ is defined as

$$H_{k,b} = \alpha_k \left[ \log_2(\frac{\alpha_k E_{k,b}}{\gamma \ln 2}) - \frac{1}{\ln 2}(1 - \frac{\gamma \ln 2}{\alpha_k E_{k,b}}) \right], \qquad \text{for } k = 1,..,K_h$$

$$H_{k,b} = \log_2(\frac{E_{k,b}}{\gamma \ln 2}) - \frac{1}{\ln 2}(1 - \frac{\gamma \ln 2}{E_{k,b}}), \qquad \text{for } k = K_h + 1,..,K$$

(17)

We conclude that, for a chosen sub-band $b$, the user with the largest $H_{k,b}$ can use the sub-band. In other words, for a sub-band $b$, if $H_{k,b}$ are different for all $k$, then

$$\rho^*_{k',b} = 1, \qquad \rho^*_{k,b} = 0 \qquad \text{for all } k \neq k' \qquad (18)$$

$$\text{where} \quad k' = \arg \max_k H_{k,b} \qquad (19)$$

## Power Allocation

Let $P^*_{k,b}$ be the optimal solution. After differentiating (13) with respect to $P_{k,b}$ by KKT optimality condition [18], we obtain

$$P^*_{k,b} = \rho_{k,b}(\frac{\alpha_k}{\gamma \ln 2} - \frac{1}{E_{k,b}}), \qquad \text{for } k = 1,..,K_h$$

$$P^*_{k,b} = \rho_{k,b}(\frac{1}{\gamma \ln 2} - \frac{1}{E_{k,b}}), \qquad \text{for } k = K_h + 1,..,K$$

(20)

As a result, in order to achieve the sub-band and power allocation, we have to compute $H_{k,b}$ and $P_{k,b}$ for all the existing UWB users. We thus need to find the set of $\alpha_k$ such that the HQoS users rate and the total power constraints are satisfied. This can be stated as

$$R'_k = \sum_{b=1}^{B} \rho_{k,b} \log_2(\frac{\alpha_k E_{k,b}}{\gamma \ln 2}) \geq R_k, \qquad \text{for } k = 1,...,K_h$$

$$\sum_{k=1}^{K} \sum_{b=1}^{B} P_{k,b} \leq P_T$$

(21)

## Interference Power Control

After allocating the sub-bands and the power to the different UWB users, we eventually need to satisfy the interference constraint. Since the interference power of a user $k$ in a sub-band $b$ depends on its allocated power in this sub-band as given by (8), we control the interference that may be caused by the UWB users to the primary users occupying the same spectrum after the power allocation. The control consists in reducing the power level of the users causing an interference level that exceeds the primary users interference threshold.

In order to be consistent with the HQoS users constraint, we have to make a certain tradeoff between the interference reduction and the QoS satisfaction. Let $P^{ic}_{k,b}$ be the power that should be allocated to the part of the sub-band $b$ causing interference. Thereby, we define a power reduction parameter $P^{red}_{k,b}$ for the different UWB users as the power that should be reduced from the allocated power in the band or the set of subcarriers that are causing interference to the primary users. In the following we define the power reduction value in the case of SQoS and HQoS users respectively.

### - For SQoS users:

For SQoS users, the allocated power is annulled in the band or the set of subcarriers occupied by a primary user. This is stated as

$$\tilde{P}^\delta_{k,b} = 0, \quad \text{and consequently} \qquad P^{red}_{k,b} = P_{k,b} \qquad (22)$$

This gives a protection to the primary user while reducing the performance of the SQoS UWB users. The performance degradation of the SQoS users should be tolerable since these users do not have strict QoS requirements.

- **For HQoS users:**

In the case of HQoS users, the interference reduction should take into consideration the rate requirements of these users. In other terms, the power is reduced to a level that can satisfy the HQoS constraints.

By taking the HQoS rate constraint given by (11), we can write

$$\sum_{b=1}^{B} \log_2(1 + P_{k,b} E_{k,b}) \;-\; \sum_{b=n_{ud}}^{n_{up}} \log_2(1 + \tilde{P}^\delta_{k,b} E_{k,b}) \geq R_k \qquad (23)$$

where $n_{up}$-$n_{ud}$ is the bandwidth occupied by a primary user. Solving (23) we obtain

$$P^{red}_{k,b} \;\leq\; \frac{2^{\frac{n_{up}-n_{ud}}{N}(r_k - R_k)} - 1}{E_{k,b}} \qquad (24)$$

This limitation in the interference reduction guarantees the QoS support of the HQoS users so that any power reduction does not affect their rate requirements.

As a conclusion, adapting the interference power reduction value to the users QoS level gives a kind of tradeoff between the need to protect the primary users and guarantee the QoS support of the HQoS UWB users.

# 5. OPTIMAL AND SUBOPTIMAL ALLOCATION ALGORITHMS

## 5.1. Mathematical characteristics of the optimal solution

To solve the formulated optimization problem, we first study the characteristics of the sub-band and power allocation functions given by (17) and (20) respectively. These two functions have the following properties:

- First, they are monotonically increasing with respect to $E_{k,b}$. This means that, for a selected sub-band, the user having better channel conditions has more chance to be assigned this sub-band with a good power level.

- Second, the two allocation functions are monotonically increasing with respect to $\alpha_k$. This can be viewed as a result of the service differentiation principle. In other terms, the functions depend on the user priority and thus, the stricter the user requirements, the higher the value of $\alpha_k$ and consequently the higher the value of these functions.

- Third, we conclude from the HQoS users constraint given by (21) that $\alpha_k$ is monotonically increasing with respect to $R_k$.

As a result, the power and the sub-band allocation functions depend on the rate constraints of the users, more precisely the HQoS users that have strict data rate requirements.

## 5.2. Optimal Algorithm

Based on the above observations, we propose an iterative algorithm detailed in Algorithm 1 for the search of the optimal sub-band and power allocation. Thereby, since the interference power control is a step that follows the power allocation step, the algorithm is divided into two parts: the joint sub-band and power allocation part, and the interference control part. In the first part, the process consists in incrementing $\alpha_k$ iteratively by a small value $\delta$ until reaching the HQoS users data rate request while respecting the power constraint. Then, using the so-obtained $\alpha_k$, the allocated power is refined in order to reduce the interference power to primary users as described in the previous section.

---

## Algorithm 1: Optimal Solution

---

### Part 1: joint sub-band and power allocation

1. Initialization
   alpha = 1
   $\alpha_k$ = alpha+$\delta$,      for $k = 1,...,k_h$
2. Sub-band allocation
   a. **for** sub-band $b = 1,...,B$
        compute $H_{k,b}$ using (17) for all $k$
        obtain $\rho_{k,b}$ and $k'$ using (18) and (19)
   b. **for** $k = 1,...,k_h$
        compute $R_k'$ using (21)
   c. **for** $k = 1,...,k_h$
        find $\hat{k}$ with $R_{\hat{k}}' < R_{\hat{k}}$ and $R_{\hat{k}}' - R_{\hat{k}} \leq R_k' - R_k$
   d. **while** $R_{\hat{k}}' < R_{\hat{k}}$
        $\alpha_{\hat{k}} = \alpha_{\hat{k}} + \delta$
        repeat a.,b. and c.
3. Power allocation
   a. compute $P_{k,b}$ using (20) for all $k$
   b. compute $P_T' = \sum_{k=1}^{K} \sum_{b=1}^{B} P_{k,b}$
   c. **if** $P_T' < P_T$
        $\alpha_k = \alpha_k + \delta/2$
      **else**
        $\alpha_k = \alpha_k - \delta/2$
      repeat 2. and 3. until $P_T' = P_T$

### Part 2: Interference power control

1. **for** $k = 1,...,K$
     compute $I_{k,b}^u$ using (8) for the allocated sub-band $b$
2. compute $I_u' = \sum_{k=1}^{K} \sum_{b=1}^{B} I_{k,b}^u$
3. **if** $I_u' > I_u^{th}$
   a. find $\hat{k}$ and the corresponding allocated sub-band $\hat{b}$ with $I_b \neq 0$
   b. **if** $\hat{k} \in \{k_h + 1,..,K\}$
        reduce the power in the subcarriers causing interference using (22)
      **else**
        reduce the power in the subcarriers causing interference using (24)

---

### 5.3. Suboptimal Algorithm

In order to reduce the high computation cost of the optimal algorithm due to the iterative process, we define a suboptimal allocation algorithm. This suboptimal algorithm is based on a cross-layer approach in the way of collecting the corresponding information from the PHY and MAC layers. The idea is first to replace $\alpha_k$ by a static parameter that can be defined once for all the users. We propose thus to use the weight parameter $W_k$ defined in section 3.1. This can be justified by the fact that the weight parameter has the same characteristics as $\alpha_k$; both parameters depend on the service data requirements or the QoS level. Second, in the suboptimal solution, the power is equally distributed among the different users and it is refined individually according to each user conditions.

The new suboptimal sub-band allocation function that will replace the optimal sub-band function given by (17) is defined as follows

$$H'_{k,b} = W_k E_{k,b} \qquad (25)$$

---

**Algorithm 2: Suboptimal Solution**

---

1. Initialization
   $$P_T^{red} = 0$$

2. Power distribution
   $$P_{k,b} = \frac{P_T}{B}, \quad \text{for } k = 1,..,K$$

3. Sub-band allocation
   **for** sub-band $b = 1,...,B$
       compute $H'_{k,b}$ using (25) for all $k$
       $k' = \arg \max H'_{k,b}$

4. Interference control
   a. **for** $k = 1,..,K$
       compute $I^u_{k,b}$ using (8) for the allocated sub-band $b$

   b. compute $I'_u = \sum_{k=1}^{K} \sum_{b=1}^{B} I^u_{k,b}$

   c. **if** $I'_u > I^{th}_u$
       find $\hat{k}$ and the corresponding allocated sub-band $\hat{b}$ with $I_{\hat{b}} \neq 0$
         **if** $\hat{k} \in \{k_h + 1,..,K\}$
         reduce the power in the subcarriers causing interference using (22)
         **else**
           reduce the power in the subcarriers causing interference using (24)
       $P'_T = P_{\hat{k},b} - P^{red}_{\hat{k},b}$
       $P^{red}_T = P^{red}_T + P^{red}_{\hat{k},b}$

5. QoS control - power refinement
   **if** $P^{red}_T \neq 0$
   a. **for** $k = 1,...,k_h$
       compute $R'_k$ using (10)
   b. **for** $k = 1,...,k_h$
       find $k'$ ($k' \neq k$) with $R'_{k'} < R_{k'}$ and $R'_{k'} - R_{k'} \leq R'_k - R_k$
       $P_{k',b} = P_{k',b} + P^{red}_T$

---

The allocation function given by (25) can be viewed as a cross-layer function since it combines the user weight as defined in (2), information provided by the MAC layer, and the user effective SINR in a sub-band as defined in (5), information provided by the PHY layer.

Algorithm 2 presents the suboptimal solution. In this algorithm, the power is first distributed equally among the existing users. Then, we proceed with the sub-band allocation using the suboptimal allocation function given by (25). The interference control is done here before the QoS satisfaction control. Thus, after the power reduction resulting from the interference power control step, the power is refined in order to ensure the unsatisfied HQoS users having no interference problem with primary users. To do so, we increase the allocated power of these unsatisfied HQoS users by the amount of the power reduced from users having interference problem with primary users.

## 6. SYSTEM PERFORMANCE

### 6.1. Channel model

The channel used in this study is the one adopted by the IEEE 802.15.3a committee for the evaluation of UWB proposals [19]. This model is a modified version of Saleh-Valenzuela model for indoor channels [20], fitting the properties of UWB channels. A log-normal distribution is used for the multipath gain magnitude. In addition, independent fading is assumed for each cluster and each ray within the cluster. The impulse response of the multipath model is given by

$$h_i(t) = G_i \sum_{z=0}^{Z_i} \sum_{p=0}^{P_i} \alpha_i(z, p)\delta(t - T_i(z) - \tau_i(z, p)) \qquad (26)$$

where $G_i$ is the log-normal shadowing of the $i$th channel realization, $T_i(z)$ the delay of cluster $z$, and $\alpha_i(z, p)$ and $\tau_i(z, p)$ represent the gain and the delay of multipath $p$ within cluster $z$, respectively.

Four different channel models (CM1 to CM4) are defined for the UWB system modelling, each with arrival rates and decay factors chosen to match different usage scenarios and to fit line-of-sight (LOS) and non-line-of-sight (NLOS) cases.

### 6.2. Simulation results

In this section, we present the simulation results for the proposed allocation scheme and we compare the performance of the optimal and the suboptimal solutions. For the simulation scenarios, we use the proposed WiMedia data rates (see Table 1) and we consider the first WiMedia channel (3.1-4.7 GHz) for CM1 channel model. Consequently, three unlicensed UWB users are considered to send simultaneously by sharing the three sub-bands of the first WiMedia channel. Moreover, we consider a licensed user occupying a bandwidth that varies from 1 to 50 MHz in the first WiMedia channel.

Since we aim at guaranteeing the QoS support of multimedia applications, we present in Fig. 2 the power satisfaction of the HQoS users causing interference on the licensed user. Two scenarios are studied: scenario 1 consists of one unlicensed HQoS UWB transmitting at a data rate of 320 Mbps and two SQoS users transmitting at 53.3 Mbps; and scenario 2 consisting of one unlicensed HQoS UWB transmitting at a data rate of 160 Mbps and two SQoS users transmitting at 53.3 Mbps. As explained before, contrarily to the SQoS users that annul their power in the primary user band, the HQoS user reduces its transmission power to a certain limit that respects its rate requirement. Accordingly, the power reduction depends on the rate requirement $R_k$ of the HQoS user and the bandwidth of the primary user occupying the same UWB user band. As shown in the figure, if the data rate requirement level of the HQoS user is high ($R_k = 320$ Mbps), the reduction is not considerable and the user satisfaction is good.

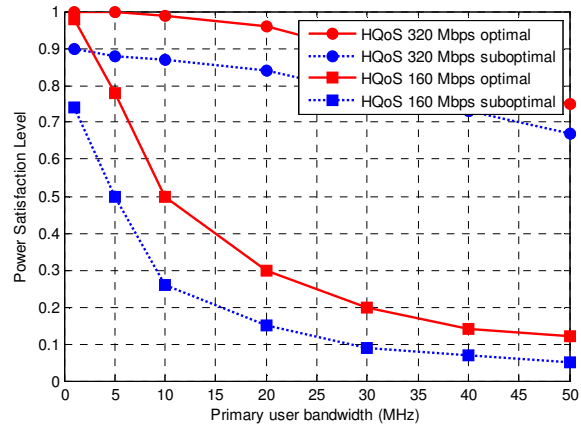

Fig. 2. Power satisfaction level of unlicensed HQoS users causing interference to a licensed user.

However, in the case where $R_k$ = 160 Mbps, we observe a substantial reduction level especially if the bandwidth occupied by the primary user is large. This proves that users with high data rate are more protected, which guarantees a good level of QoS support for multimedia applications having high rate requirements. Besides, we notice that the optimal solution outperforms the suboptimal solution with an average of 10% in both scenarios.

In Fig. 3, we present the interference reduction ratio in the cross-layer (suboptimal) solution for the HQoS and the SQoS users in different conditions. We mean by interference reduction ratio the ratio of the whole reduced interference level resulting from the power refinement of the users causing interference to primary users, to the original interference level obtained before applying the new allocation scheme. We present results according to three values of $I_{th}$ as shown in the figure. We consider a scenario which consists of one unlicensed HQoS UWB transmitting at a data rate of 320 Mbps and two SQoS users transmitting at 53.3 Mbps. As we can observe, increasing the value of $I_{th}$ decreases the interference reduction level, which is normal since it is a constraint imposed by the primary user to the UWB users. Besides, note that the interference reduction in the case of SQoS users is more important. This is resulted from the fact that the interference depends on the power which is annulled in the part of the band causing interference by SQoS users on a primary user whereas it is reduced in the case of HQoS users.

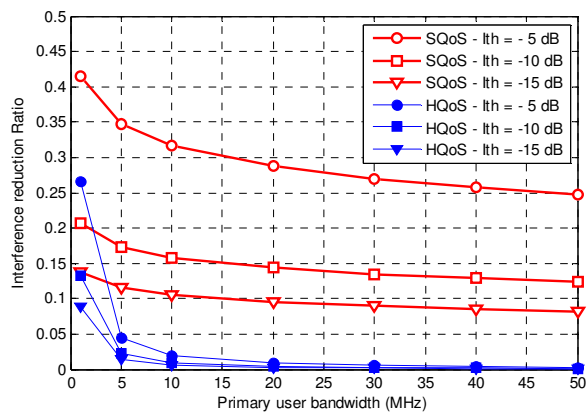

Fig. 3. Interference reduction ratio for the different users in different conditions

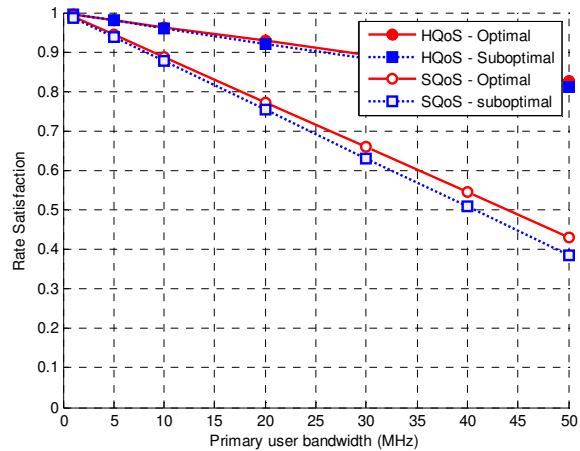

.

Fig. 4. Rate Satisfaction of HQos and SQoS users in the proposed scheme.

In Fig. 4, we present the average rate satisfaction level of all the users in the optimal and suboptimal solutions. We consider the same scenarios defined for Fig. 2. As shown in the figure, HQoS users rate satisfaction level is much more better than that of SQoS users. This proves that the QoS support of the HQoS users is guaranteed in terms of rate satisfaction in both optimal and suboptimal solutions. For instance, in the case where an HQoS user is causing interference to a primary user occupying a bandwidth of 50 MHz, the rate satisfaction level is almost 80%. On the other hand, since the SQoS users have less QoS requirements, their rate satisfaction level decreases to 40% when they cause interference to a primary user occupying a bandwidth of 50 MHz in the suboptimal solution. Besides, we note that the optimal and suboptimal solutions perform very close in both HQoS and SQoS cases. This can be justified by the fact that, in both solutions, the sub-band allocation is the same, so the channel quality is the same and what differs is the assigned power level.

## 7. CONCLUSION

In this paper, we have studied the resource allocation problem for the unlicensed high-rate UWB systems which use an underlay approach for the coexistence with already existing licensed systems. This study gives an answer to the spectrum sharing problem for the secondary users having heterogeneous conditions and aiming at coexisting with the primary users.

The problem has been first studied analytically by deriving a constrained optimization problem. This study has lead to an optimal solution while considering three main constraints: the QoS requirements, the channel quality, and the interference level caused by the UWB users on the primary users. A suboptimal solution is also proposed to reduce the complexity of the optimal solution. It is based on a cross-layer approach in the way it jointly considers information provided by the PHY and MAC layers.

Finally, we have shown through simulations the efficiency of the proposed allocation scheme that guarantees a good QoS support for users with strict requirements. Besides, the slight performance degradation of the unlicensed users that do not have QoS requirements is viewed as a sacrifice to ensure an efficient use of the spectrum that limits the interference affecting the primary users. We have shown also that the optimal and suboptimal solutions perform close, which means that the new simplified cross-layer approach is advantageous and can be an efficient solution for the QoS support and spectrum sharing matters in the next generation UWB systems.

## ACKNOWLEDGEMENTS

The research leading to these results has received funding from the European Community's Seventh Framework Programme FP7/2007-2013 under grant agreement n° 213311 also referred as OMEGA.

## REFERENCES

[1]     J. Mitola and G. Maguire, "Cognitive Radio: Making software radio more personal," *IEEE Journal on Personal Communications*, vol. 6, no. 4, pp. 13-18, Aug. 1999.

[2]     "First report and order, revision of part 15 of the commission's rules regarding ultra-wideband transmission systems," FCC, ET Docket 98-153, Feb. 14, 2002.

[3]     WiMedia Alliance, Inc., "Multi-band OFDM physical layer specification," Release 1.1, July 2005.

[4]     Standard ECMA-368, High rate ultra wideband PHY and MAC standard, 2$^{nd}$ edition, Sept. 2007.

[5]     C.Y. Wong, R.S. Cheng, K.B. Lataief and R.D. Murch, "Multiuser OFDM with adaptive subcarrier, bit and power allocation," *IEEE Journal on Selected Areas in Communications*, vol. 17, no. 10, pp. 1747-1758, Aug. 2002..

[6]     I. Kim, I. Park and Y.H. Lee, "On the use of linear programming for dynamic subcarrier and bit allocation in multiuser OFDM," *IEEE Trans. on Vehicular Technology*, vol. 55, pp. 1195-1207, July 2006.

[7]     A. Attar, O. Holland, M.R. Nakhai and A.H. Aghvami, "Interference-limited resource allocation for cognitive radio in orthogonal frequency-division multiplexing networks*," IET Communications*, vol. 2, no. 6, pp. 815-816, July 2008.

[8]     T. Qin and C. Leung, "A cost minimization algorithm for a multiuser OFDM cognitive radio system," *IEEE Pacific Rim Conference on Communications, Computers and Signal* (*PacRim'07*), pp. 518-521, Aug. 2007.

[9]     G. Bansal, J. Hossain and V.K. Bhargava, "Optimal and suboptimal power allocation schemes for OFDM-based cognitive radio systems," *IEEE Trans. on Wireless Communications*, vol. 7, no. 11, pp. 4710-4718, Nov. 2008.

[10]    Z. Chen, D. Wang and G. Ding, "An OFDM-UWB scheme with adaptive carrier selection and power allocation," *IEEE Intern. Conference on Wireless Communications, Networking and Mobile Computing (WiCOM'06)*, pp.1-4, China, Sept. 2006.

[11]    W.P. Siriwongpairat, Z. Han and K.J. Ray Liu, "Power controlled channel allocation for multi-user multiband UWB systems," *IEEE Trans. on Wireless Communications,* vol. 6, no. 2, pp. 583–592, Feb. 2007.

[12]    B. Muquet, Z. Wang, G.B. Giannakis, M. de Courville and P. Duhamel, "Cylic pefix or zero padding for wireless multicarrier transmission," *IEEE Trans. on Communications*, vol. 50, pp. 2136-2148, Dec. 2002.

[13]    3GPP TSG-RAN-1, "R1-030999: considerations on the system performance evaluation of HSDP using OFDM modulations"; RAN WG1 #34.

[14]    3GPP TSG-RAN-1, "R1-040090: system level performance evaluation for OFDM and WCDMA in UTRAN," Finland, Jan. 2004.

[15]    T. Weiss, J. Hillenbrand, A. Krohn and F.K. Jondral, "Mutual interference in OFDM-based spectrum pooling systems," *IEEE Vehicular Technology Conference (VTC'04-spring)*, pp. 1873-1877, May 2004.

[16]    S. Maheshwari, "An Efficient QoS scheduling architecture for IEEE 802.16 wireless MANs," Indian Institute of Technology, India, 2005.

[17]    S. Boyd and L. Vandenberghe, *Convex Optimization*. Cambridge University Press, 2004.

[18]    D.P. Bertsekas, *Nonlinear Programming 2$^{nd}$ Edition*. Athena Scientific, 1999.

[19]    J. Foester, "Channel Modeling sub-committee report (final)," IEEE P802.15.-02/490rl-SG3a, 2003.

[20]    A. Saleh and R. Valenzuela, "A statistical model for indoor multipath propagation," *IEEE Journal on Selected Areas in Communications*, vol. 5, no. 2, pp. 128–137, Feb. 1987.

**Authors**

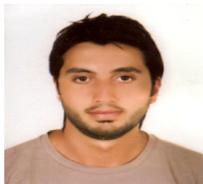

**Ayman Khalil** received the M.Sc in Networking and Telecommunications from the Lebanese University/Saint Joseph University, Beirut, Lebanon in 2007. Actually, he is working on a Ph.D. study in the Institute of Electronics and Telecommunications of Rennes (IETR), Rennes, France. His research interests lie in wireless communication systems, and particularly focus on topics related to all aspects of PHY layer, MAC layer and cross-layer designs. He is involved in the European OMGA project and his main contribution focuses on studies and proposals holding solutions for the cross-layer resource allocation and spectrum sharing matters for the next generation home networks.

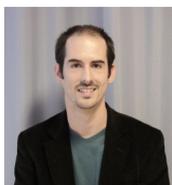

**Matthieu Crussiere** received the M.Sc. and Ph.D. degrees in electrical engineering from the National Institute of Applied Sciences (INSA), France, in 2002 and 2005, respectively. During its Ph.D. he was with the Electronics and Telecommunications Institute of Rennes (IETR), where he worked on the optimization of high-bit rate powerline communications. Since 2005, he has been an Associate Professor in the Department of Telecommunications and Electronic Engineering at INSA and currently leads its research activities at IETR. His main research interests lie in digital communications and signal processing techniques, and particularly focus on multi-carrier spread-spectrum systems, synchronization, channel estimation, and adaptive resource allocation. He has been involved in several European and national research projects including powerline communications, broadcasting systems, ultra wideband and mobile radio communications.

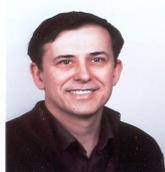

**Professor Jean-François Hélard** received his Dipl.-Ing. and his Ph.D in electronics and signal processing from the National Institute of Applied Sciences (INSA) in Rennes, France, in 1981 and 1992, respectively. From 1982 to 1997, he was research engineer and then head of channel coding for the digital broadcasting research group at CCETT (France Telecom Research Center) in Rennes. In 1997, he joined INSA, where he is currently Professor and Deputy Director of the Rennes Institute for Electronics and Telecommunications (IETR), created in 2002 in association with the CNRS. His research interests lie in signal processing techniques for digital communications, such as space-time and channel coding, multi-carrier modulation, as well as spread-spectrum and multiuser communications. He is involved in several European and national research projects in the fields of digital video terrestrial broadcasting, mobile radio communications and cellular networks, power-line and ultra-wide-band communications. Prof. J-F. Hélard is a senior member of IEEE, author and co-author of more than 130 technical papers in international scientific journals and conferences, and holds 13 European patents.